\documentclass[conference]{IEEEtran}
\IEEEoverridecommandlockouts
\usepackage{cite}
\usepackage{amsmath,amssymb,amsfonts}
\usepackage{algorithmic}
\usepackage{graphicx}
\usepackage{textcomp}
\usepackage{xcolor}
\usepackage{threeparttable}
\usepackage{booktabs}
\usepackage{multirow}
\usepackage{subfigure}
\usepackage{colortbl}
\usepackage[table]{xcolor}
\usepackage{url}
\def\BibTeX{{\rm B\kern-.05em{\sc i\kern-.025em b}\kern-.08em
    T\kern-.1667em\lower.7ex\hbox{E}\kern-.125emX}}
\begin{document}

\title{Adversarial Attacks and Robust Defenses in Speaker Embedding based Zero-Shot Text-to-Speech System \\
\thanks{* Corresponding Author: Ming Li}
}
% \author{Anonymous}
\author{
    \IEEEauthorblockN{Ze Li$^{1,2}$, Yao Shi$^{3}$, Yunfei Xu$^3$,  Ming Li$^{1,2*}$ }
    \IEEEauthorblockA{$^1$ School of Computer Science, Wuhan University, Wuhan, China}
    \IEEEauthorblockA{$^2$ Suzhou Municipal Key Laboratory of  Multimodal Intelligent Systems, Digital Innovation Research Center, \\ Duke Kunshan University, Kunshan, China}
    \IEEEauthorblockA{$^3$ AI Center, OPPO, Beijing, China}
    {ming.li369@dukekunshan.edu.cn}
}

\maketitle

\begin{abstract}
Speaker embedding based zero-shot Text-to-Speech (TTS) systems enable high-quality speech synthesis for unseen speakers using minimal data. However, these systems are vulnerable to adversarial attacks, where an attacker introduces imperceptible perturbations to the original speaker’s audio waveform, leading to synthesized speech sounds like another person. This vulnerability poses significant security risks, including speaker identity spoofing and unauthorized voice manipulation. This paper investigates two primary defense strategies to address these threats: adversarial training and adversarial purification. Adversarial training enhances the model's robustness by integrating adversarial examples during the training process, thereby improving resistance to such attacks. Adversarial purification, on the other hand, employs diffusion probabilistic models to revert adversarially perturbed audio to its clean form. Experimental results demonstrate that these defense mechanisms can significantly reduce the impact of adversarial perturbations, enhancing the security and reliability of speaker embedding based zero-shot TTS systems in adversarial environments.
\end{abstract}


\section{Introduction}
\label{sec:intro}
\begin{figure}
  \centering
      \subfigure[Traditional speaker embedding based zero-shot TTS system.]{
        \includegraphics[width=\linewidth]{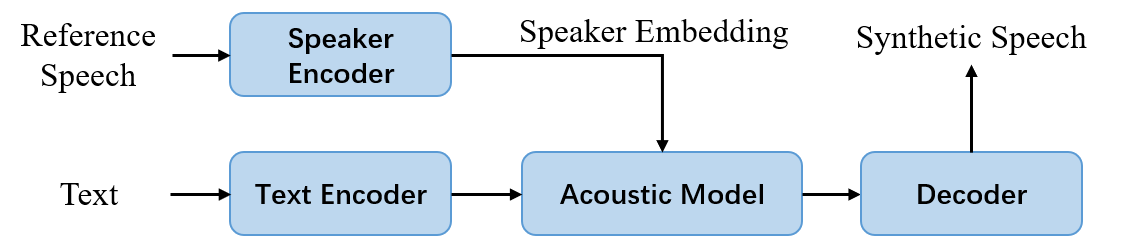}
        \label{fig:figure1}
    }
    \subfigure[LLM-based speaker embedding based zero-shot TTS system.]{
        \includegraphics[width=\linewidth]{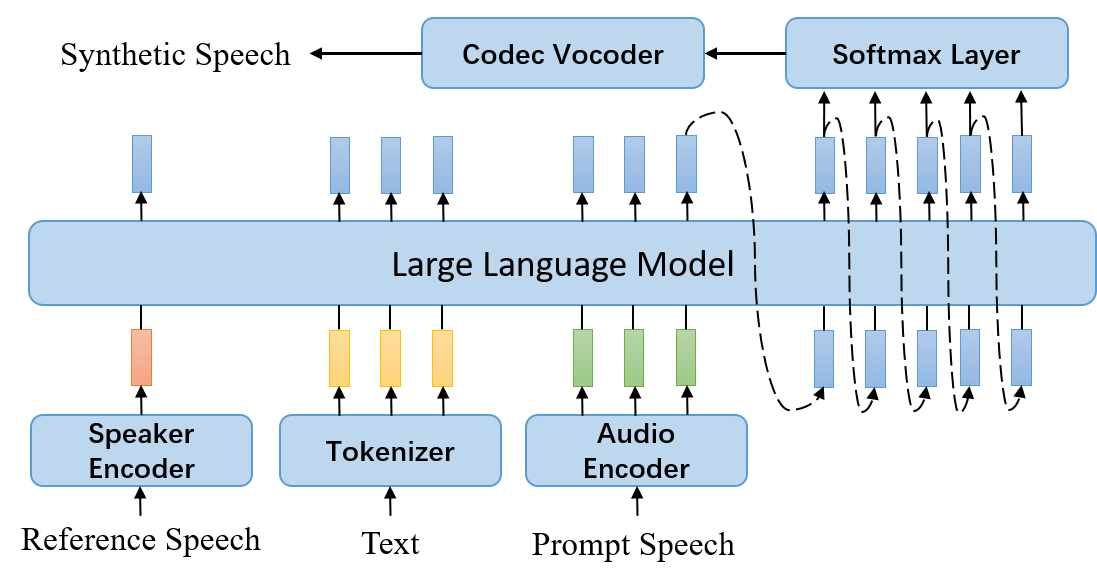}
        \label{fig:figure2}
    }
  \caption{Architecture of traditional and LLM-based speaker embedding based zero-shot TTS systems.}
  \label{fig:zs-tts}
\end{figure}

With the rapid advancement of deep learning technologies, Text-to-Speech (TTS) systems have made significant progress \cite{fastspeech2, glowtts, vits}, particularly with the emergence of Zero-Shot TTS. This technology enables the generation of natural speech for any speaker from short audio samples. Currently, the mainstream Zero-Shot TTS approaches include speaker embedding-based methods \cite{embedding-based-1, embedding-based-2, embedding-based-3}, the large language model (LLM) based ones \cite{vall-e-1, vall-e-2} and their combinations. As shown in Fig.\ref{fig:zs-tts}, speaker embedding based Zero-Shot TTS utilizes a speaker encoder alongside a TTS component, while the large speech generation models formulate the Zero-Shot TTS task as a language modeling task within the neural codec domain.

Although Zero-Shot TTS technology has shown great potential, it also faces new challenges, particularly regarding security and robustness. Malicious attacks have become a severe concern as these systems are increasingly deployed in scenarios that demand high security and reliability. In particular, speaker embedding based Zero-Shot TTS systems are vulnerable to various spoofing attacks. Research works have shown that even with the deep neural networks based approaches \cite{speaker-encoder, x-vector, epaca-tdnn}, speaker encoders are susceptible to malicious spoofing attacks such as impersonation \cite{impersonation}, replay attacks \cite{replay}, voice conversion \cite{vc}, and adversarial attacks \cite{adv}. 
Attackers can manipulate their own voice to alter the extracted speaker embeddings, leading the system to generate speech resembling the target speaker, thereby enabling a range of fraudulent activities, including voice forgery and identity impersonation.

\begin{figure*}
  \centering
  \includegraphics[width=\linewidth]{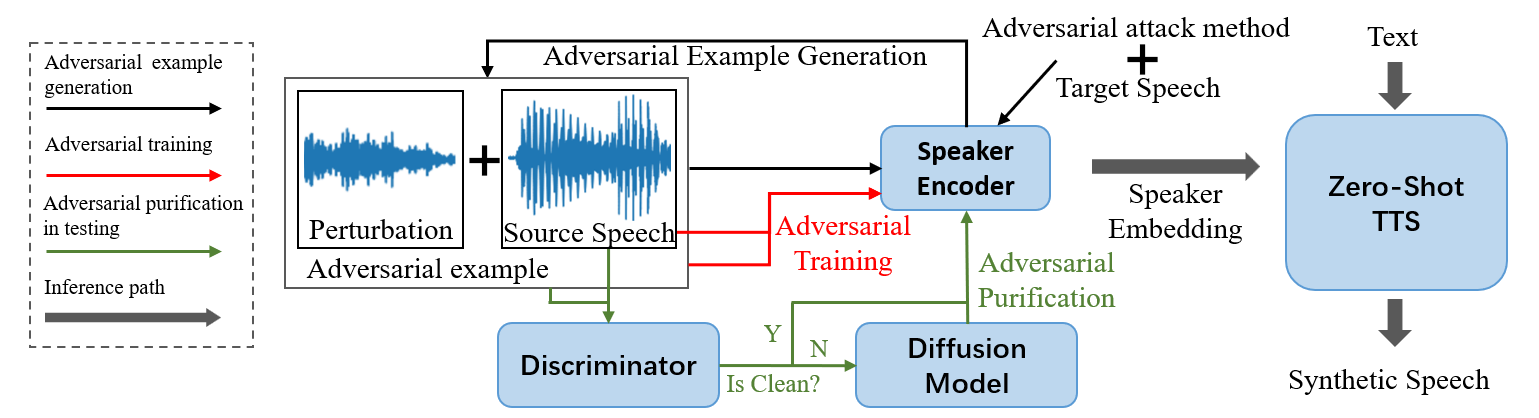}
  \caption{Attack and defense framework for speaker embedding based zero-shot TTS system.}
  \label{fig:system_framework}
\end{figure*}

This study focuses on the adversarial attacks in speaker embedding based Zero-Shot TTS systems. Adversarial attacks are typically carried out by generating adversarial examples, which are crafted by introducing imperceptible perturbations through optimization methods such as Fast Gradient-Sign Method \cite{fgsm}, Projected Gradient Descent (PGD) \cite{pgd}, optimizer-based \cite{optimizer-based}, and Carlini-Wagner \cite{cw} to lead a misclassification.

The prevailing method for defending against adversarial attacks is adversarial training \cite{adv-train1, adv-train2}, which enhances the model’s robustness by exposing it to adversarial examples during the training phase. Although adversarial training is widely regarded as the most effective defense strategy, it requires substantial computational resources, and the model remains susceptible to unseen attacks that differ from the adversarial methods used during training. Another approach is adversarial purification, which focuses on designing effective purification models to mitigate the adversarial perturbations in input samples. Currently, diffusion models have proven to be the state-of-the-art purification models in both the vision domain \cite{diffusion_survey} and the audio domain, such as in background noise removal \cite{diff_sv}, speaker verification \cite{sv_diff}  and speech command recognition \cite{diff_sc} tasks. 

This paper investigates adversarial attacks and defenses for speaker embedding based Zero-Shot TTS systems. We experiment with both traditional and LLM-based speaker embedding based zero-shot TTS systems. For the attack phase, we utilize PGD \cite{pgd} and Adam \cite{adam} optimizer-based approaches to generate adversarial examples targeting the speaker encoder of the Zero-Shot TTS system in a white-box attack scenario. For the defense phase, we evaluate and compare two strategies: an active defense through adversarial training and a passive defense via adversarial purification using diffusion models.
The Demo Page can be found here\footnote{https://se-zs-tts.github.io/}.

\section{Methods}
This section presents our methods for adversarial attacks on speaker embedding based zero-shot TTS system, along with corresponding defense measures, including adversarial training and adversarial purification. The overall framework is illustrated in Fig. \ref{fig:system_framework}.

\subsection{Adversarial Example Generation} \label{adv_gen}
Adversarial examples refer to instances with imperceptible perturbations that are deliberately introduced. These perturbations are obtained by solving an optimization problem and lead a well-trained model to make incorrect predictions. This study utilizes PGD \cite{pgd} and Adam \cite{adam} optimizer-based methods to generate adversarial examples. Both methods are gradient-based white-box attacks. We aim to attack the speaker encoder of a speaker embedding based Zero-Shot TTS system by adding small perturbations to the input speech, causing the output speaker embeddings to change, which leads the TTS system to generate speech that mimics the target speaker.

The core idea of the PGD method is to iteratively update the input samples based on the gradient sign of the loss function while constraining the perturbation magnitude. Specifically, given input speeches $x$, target labels $y'$ and a well-trained speaker encoder model $f(\cdot)$. The predicted labels $\hat{y}$ can be obtained by computing the index of the maximum cosine similarity between the speaker embeddings of the source speeches and the spoofed ones with adversarial examples, which are extracted by the speaker encoder. The adversarial perturbations $\delta$ can be generated by:
\begin{equation}
    \hat{y} = \left[ \arg \max_j cosine(f(x)_{i}, f(x+\delta)_{j} ) \right]_{i} \label{y_hat}
\end{equation}

\begin{equation}
\begin{aligned}
    \delta_{t+1} = \delta_{t} -  \alpha \cdot sign&(\nabla_{\delta_{t}}Loss(\hat{y}, y')), \\
    \text{s.t.} \quad \Vert \delta & \Vert_{\infty} \leq \epsilon  \label{pgd}
\end{aligned}
\end{equation}
where $sign(\cdot)$ represents the sign of the gradient, $\alpha$ and $\epsilon$ control the magnitude of each update and the maximum allowable perturbation, respectively.

Compared to PGD, Adam is an adaptive learning rate optimization algorithm. In each iteration, Adam not only utilizes the current gradient information but also incorporates momentum from previous iterations to update the perturbation:

\begin{equation}
\begin{aligned}
     m_t &= \beta_1 m_{t-1} + (1 - \beta_1) \nabla_{\delta_t}Loss(\hat{y}, y'), \\
    v_t &= \beta_2 v_{t-1} + (1 - \beta_2) (\nabla_{\delta_t}Loss(\hat{y}, y'))^2, \\
     \delta_{t+1} &= \delta_t - lr \cdot \frac{m_t}{\sqrt{v_t} + \xi}, \quad \text{s.t.} \quad \Vert \delta \Vert_{\infty} \leq \epsilon \label{adam}
\end{aligned}
\end{equation}
where $m_t$ and $v_t$ are the second-moment and second-moment estimates of the gradient, respectively. $lr$ is the learning rate, $\xi$ is the numerical stability constant, $\beta_{1}$ and $\beta_{2}$ are the decay rates for the first and second-moment estimates, respectively.

\subsection{Adversarial training}
Adversarial training is one of the most widely used and effective methods for defending against adversarial attacks, as it enhances the model's robustness by incorporating adversarial examples into the training process. In this work, we also employ adversarial training. For each speech sample within a batch $\{(x_i, y_i)\}_{i=1}^b$, we randomly assign a target speaker label $y'$ that differs from the source speaker and then apply adversarial attack methods to generate adversarial examples $\{(\hat{x}_i, y'_i)\}_{i=1}^b$. Subsequently, these adversarial examples are labeled with the source speaker's label and are used alongside the source speech $\{(\hat{x}_i, y_i) \cup (x_i, y_i)\}_{i=1}^b$ to fine-tune the well-trained speaker encoder model. Finally, the fine-tuned speaker encoder model is used to retrain the zero-shot TTS system.

\subsection{Adversarial Purification}
Diffusion-based adversarial purification is an emerging defense technique against adversarial attacks, which utilizes diffusion models to remove adversarial perturbations from input data, thereby restoring clean speech for effective defense. 
As a plug-and-play module, diffusion models effectively circumvent the issues of domain shifts and secondary training associated with adversarial training. Furthermore, they do not require training on predefined adversarial examples, which endows them with solid generalization capabilities and allows them to address a wide range of attack methods.

A diffusion model normally consists of a forward diffusion process and a reverse sampling process. The forward diffusion process gradually adds Gaussian noise to the input speech until the distribution of the noisy speech converges to a standard Gaussian distribution:
\begin{equation}
    q(x_t \mid x_0) = \mathcal{N}(x_t; \sqrt{\bar{\alpha}_t} x_0, (1 - \bar{\alpha}_t) \mathbf{I})
\end{equation}
where $x_0$ is the clean speech, $x_t$ represents the noisy speech at time step $t$, hyperparameter $\bar{\alpha}_t$ controls the noise level. 

The reverse sampling process takes the standard Gaussian noise as input and gradually denoises the noisy speech to recover clean speech. The reverse process is approximated by learning a model $p_\theta(x_{t-1} \mid x_t)$:
\begin{equation}
    p_\theta(x_{t-1} \mid x_t) = \mathcal{N}(x_{t-1}; \mu_\theta(x_t, t), \Sigma_\theta(x_t, t))
\end{equation}
where $\mu_\theta(x_t, t)$ and $\Sigma_\theta(x_t, t)$ are represent the predicted mean and covariance for time step $t$, respectively.

The optimization objective is to minimize the speech reconstruction error, which is achieved using a Mean Squared Error (MSE) loss function:
\begin{equation}
    L = MSE(x_0, \hat{x_0}) = \Vert x - \hat{x_0} \Vert_2^2 / N
\end{equation}
where $\hat{x_0}$ represents the speech obtained by denoising the noisy speech $x_t$, $N$ is the number of samples in speech $x_0$.

Additionally, considering that adversarial purification might affect clean audio, we introduce a binary classifier before the diffusion module to distinguish between audio samples with adversarial perturbations and those that are clean.

\renewcommand{\arraystretch}{1.0}
\begin{table*}[htbp]\centering 
    \scriptsize %\footnotesize \scriptsize
    \caption{The performance of speaker embedding based zero-shot TTS systems under various defense modes against different attack methods. Ori., Tgt., Adv., and Adv.(Syn) represent the source speech, target speech, adversarial samples, and the speech synthesized by the TTS system from the adversarial samples, respectively. Model A and B represent the traditional and LLM-based speaker embedding based zero-shot TTS systems, respectively.}
    \tabcolsep=0.4em
     \label{tab:res}
\begin{threeparttable}
    \begin{tabular}{cccccccccc}
    \toprule

    \multirow{2}*{\textbf{Model}} &\multirow{2}*{\textbf{Defense}} &\multirow{2}*{\textbf{Attack Method}} &\textbf{Attack Success} &\textbf{Defense Success} &\textbf{Ori. vs Adv.} &\textbf{Tgt. vs Adv.} &\multirow{2}*{\textbf{EER[\%]}} &\textbf{Ori. vs Adv.(Syn)} &\textbf{Tgt. vs Adv.(Syn)}   \\
    & & &\textbf{Rate[\%]} &\textbf{Rate[\%]} &\textbf{Similarity} &\textbf{Similarity} &  &\textbf{Similarity} &\textbf{Similarity} \\
    \midrule
   \multirow{9}*{\normalsize A} & & None & - & - & - & - & 0.957 & 0.370 & -0.001 \\
    & & Adam-based & 99.53 & 0.47 & 0.134 & 0.934 & 0.957 & 0.048 & 0.291 \\
     & \multirow{-3}*{None} & PGD & 100 & 0 & 0.074 & 0.959 & 0.957 & 0.023 & 0.311 \\
     \cmidrule(lr){2-10} 
   & \multirow{2}*{\parbox{3cm}{\centering Adversarial Training \\ with Adam-based Attack}} & Adam-based & 9.65 & 90.35 & 0.747 & 0.421 & 2.350 & 0.296  & 0.149 \\
    & & PGD & 37.85 & 62.15 & 0.618 & 0.546 & 2.350 & 0.239 & 0.190 \\
     \cmidrule(lr){2-10}
    & & Adam-based & 1.56 & 98.44 & 0.839 & 0.335 & 4.626 & 0.364 & 0.126 \\
   & \multirow{-2}*{\parbox{3cm}{\centering Adversarial Training \\ with PGD Attack}} & PGD & 4.02 & 95.98 & 0.781 & 0.392 & 4.626 & 0.331 & 0.145 \\
   \cmidrule(lr){2-10}
   & \multirow{2}*{Adversarial Purification} & Adam-based & 0.39 & 91.41 & 0.549 & 0.183 & 0.957 & 0.181 & 0.048 \\
    & & PGD & 2.34 & 83.98 & 0.479 & 0.157 & 0.957 & 0.154 & 0.044\\
    \midrule
   \multirow{5}*{\normalsize B} & & None & - & - & - & - & 0.484 & 0.568 & 0.190 \\
    & & Adam-based & 99.22 & 0.78 & 0.244 & 0.759 & 0.484 & 0.241 & 0.339 \\
     & \multirow{-3}*{None} & PGD & 100 & 0 & 0.157 & 0.841 & 0.484  & 0.201 & 0.377 \\
     \cmidrule(lr){2-10} 
   & \multirow{2}*{Adversarial Purification} & Adam-based & 0.78 & 89.1 & 0.587 & 0.273 & 0.484 & 0.376 & 0.212 \\
    & & PGD & 0 & 92.2 & 0.619 & 0.238 & 0.484 & 0.411 & 0.201 \\
    \bottomrule
    \end{tabular}
    \begin{tablenotes}
        \item *When both defense and attack are set to None, Adv.(Syn) refers to the speech synthesized by the TTS system from the source speech.
    \end{tablenotes}
\end{threeparttable}
\end{table*}

\section{experimental settings}
\subsection{Speaker Embedding Baesd Zero-Shot TTS Training}

\subsubsection{Speaker Encoder Training}

We utilize the ResNet \cite{resnet34} architecture as the speaker encoder model, including the ResNet34-based and ResNet101-based ones. The residual block channels are set to \{64,128,256,512\}, and the output feature maps are aggregated with a global statistics pooling layer that calculates each feature map's means and standard deviations. The acoustic features are 80-dimensional log Mel-filterbank energies with a frame length of 25ms and a hop size of 10ms.

The speaker encoder model is pretrained on the VoxCeleb2 \cite{vox2dev} development set and tested on the VoxCeleb1-O \cite{voxceleb1} test set. We adopt the on-the-fly data augmentation \cite{data_aug} to add additive background noise or convolutional reverberation noise for the time-domain waveform. The MUSAN \cite{musan} and RIR Noise \cite{RIR} datasets are used as noise sources and room impulse response functions, respectively. The speed perturbation \cite{dku_voxsrc20}, which speeds up or down each utterance by a factor of 0.9 or 1.1, is applied to yield shifted pitch utterances that are considered from new speakers. The input utterances are truncated to 2 seconds. We employ the ArcFace \cite{arcface} classifier, with the margin and scale parameters set as 0.2 and 32, respectively. Network parameters are updated using an SGD optimizer with an initial learning rate of 0.1. The learning rate is decayed by a factor of 0.1 every 10 epochs until 1e-5.

\subsubsection{Traditional Zero-Shot TTS system Training}
We utilize the VITS \cite{vits} structure as the traditional zero-shot TTS component. The speaker encoder is the ResNet34-based one. We use the clean subsets of the train and development sets from LibriTTS \cite{libritts} to train the zero-shot TTS component. The speaker embeddings for the utterances are obtained from the well-trained speaker encoder. The TTS network parameters are updated using the AdamW \cite{adamw} optimizer with a learning rate 2e-4. The batch size is 32, and the total epoch is 40.

\subsubsection{LLM-based Zero-Shot TTS system Training}
The LLM-based zero-shot TTS system is built on the LauraGPT model \cite{lauragpt}. The speaker encoder is the ResNet101-based one. For audio codec, we utilize a pre-trained open-source codec model from the Funcodec toolkit \cite{funcodec}, and the text tokenizer is sourced from Qwen \cite{qwen}. We train the system using the WenetSpeech4TTS \cite{wenetspeech4tts} training set and HQ-Conversations \cite{iscslp} dataset. The training process consists of two stages. In the first stage, we pre-trained the system using the Premium subset of WenetSpeech4TTS, which includes 945 hours of speech data. The learning rate is set to 1e-3, with 10,000 warm-up steps, a batch size of 160, and a total of 50 epochs. In the second stage, we fine-tuned the system on the 100-hour HQ-Conversations dataset, setting the learning rate to 1e-4, the batch size to 160 and completing 360 epochs.

\subsection{Speaker Encoder Adversarial Training}
The adversarial attack methods are described in \ref{adv_gen}. For the PGD method, the perturbation limit $\epsilon$ is set to the 5\% of the maximum amplitude of each audio sample, with 20 iterations and the step size $\alpha$ is decreased from 4e-3 to 4e-4 with cosine delay. For the Adam optimizer-based method, $\epsilon$ is set to the 5\% of the maximum amplitude of each audio sample, with 50 iterations and the learning rate $lr$ that decays from 1e-3 to 1e-5 with cosine delay.
To balance training costs, we adopt a relatively large perturbation range to accelerate adversarial sample generation. However, one can reduce the upper bound of $\epsilon$ and increase the number of iterations to obtain perturbations that are more imperceptible.

In adversarial training, the batch-wise random targeted attack strategy is employed, where a batch of speech samples is selected, and each sample within the batch is randomly assigned a target sample with a different speaker identity. Adversarial examples are then generated using the adversarial attack method and combined with the original speeches to fine-tune the speaker encoder model. No data augmentation is applied. The model is optimized using an Adam optimizer with a cosine decay learning rate schedule, starting at 1e-3 and decaying to 1e-5. The batch size is 256, and the number of epochs is 3.

\subsection{Diffusion-based Adversarial Purification Training}
DiffWave \cite{diffwave}, a representative diffusion model in the waveform domain, is used as a defensive purification model. We use the same settings as those in \cite{diff_sc} for diffusion parameters. The VoxCeleb2 development set is employed for model training, with input utterances truncated to 2 seconds. The learning rate is 2e-4, and the batch size is 16.

We introduce a ResNet18-based binary classifier before the diffusion model to prevent the diffusion model from damaging normal speech. The ArcFace (m=0.2, s=32) classifier is introduced to identification. A subset of the VoxCeleb2 development set is selected, from which 256,000 adversarial samples are generated using adversarial attack methods and split 9:1 for training and testing. The model is updated using the Adam optimizer with a cosine decay learning rate schedule, starting at 1e-3 and decaying to 1e-5. The batch size is 256, and the total number of epochs is 10.

\section{RESULTS AND ANALYSIS}
We used the adversarial attack methods described in \ref{adv_gen} to randomly generate 2,560 adversarial samples for each method in the VoxCeleb2 data set and the WenetSpeech4TTS development set, and evaluated the performance of both the traditional and LLM-based zero-shot TTS systems. Table \ref{tab:res} presents the results for each method in terms of attack efficacy, defense performance, and synthesis quality. We define the defense as successful if the speaker embedding of the adversarial sample is most similar to the source speech's. Conversely, the attack is successful if the adversarial sample's speaker embedding is most similar to the target speech's.

\subsection{Adversarial Attack Results}
Both attack methods exhibited substantial effectiveness, nearly achieving a 100\% attack success rate. In model A, after applying the attacks, the cosine similarity between the adversarial samples and the target speech was measured at 0.934 and 0.959, respectively. In contrast, the cosine similarity between the adversarial samples and the source speech decreased to 0.134 and 0.074. Quantitatively, the PGD attack method demonstrated greater effectiveness compared to the Adam-based attack method. The same conclusion can be drawn for Model B.

\subsection{Adversarial Training Defense Results}
After incorporating adversarial samples into the training process, the model's robustness improved, with defense success rates rising from 0.47\% and 0\% to 90.35\% and 95.98\%, respectively. However, the introduction of adversarial samples, as cross-domain data, caused a degree of performance degradation on normal data. The impact of this degradation was more pronounced with stronger attack methods, as evidenced by the EER on the Vox1-O, which increased from 0.957\% to 2.35\% and 4.626\%. Additionally, it is noteworthy that models trained with adversarial samples generated by the weaker Adam-based attack exhibited significantly reduced defense performance against stronger PGD attacks. In contrast, models trained with PGD-generated adversarial samples retained strong defense capabilities against Adam-based attacks, achieving a high defense success rate of 98.44\%.

\begin{figure}
  \centering
      \subfigure[Attack and defense success rate across different diffusion steps.]{
        \includegraphics[width=0.95\linewidth]{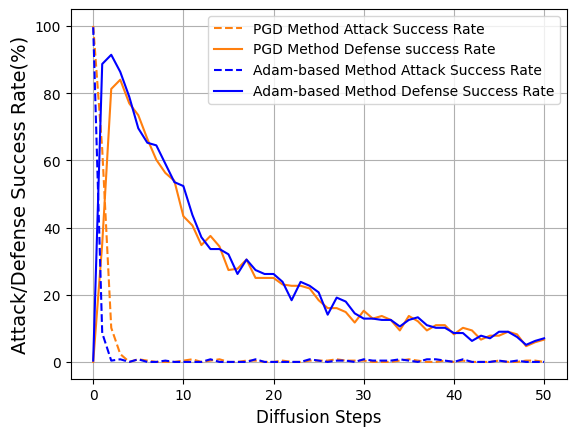}
        \label{fig:suc}
    }
    \subfigure[The similarity of speaker embeddings between the adversarially purified speech and the source and target speech across different diffusion steps.]{
        \includegraphics[width=0.95\linewidth]{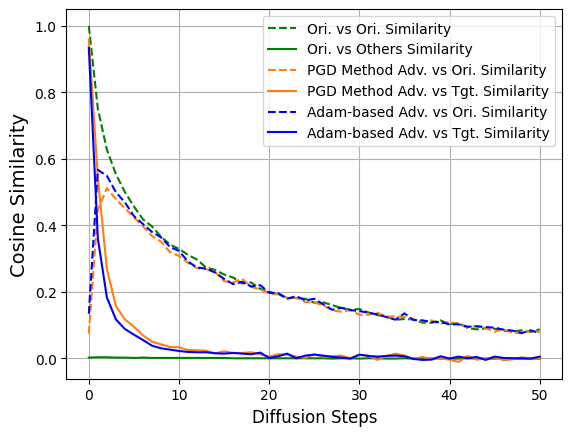}
        \label{fig:sim}
    }
  \caption{Adversarial purification performance across different diffusion steps.}
  \label{fig:suc_sim}
\end{figure}

\subsection{Adversarial Purification Defense Results}
Diffusion-based Adversarial Purification is indeed a promising emerging technology. 
It can serve as a pre-processing module, providing a robust defense against various attack methods without re-training the speaker encoder and the TTS system.
In model A, we observe that after adversarial purification, the attack success rates of adversarial samples decreased from 99.53\% and 100\% to 0.39\% and 2.34\%, respectively. However, the defense success rates improved by only 91.41\% and 83.98\%. This is because the perturbations removed by the diffusion module cannot perfectly match the added perturbations. Consequently, the denoised speaker embedding may resemble those of other speakers rather than the source speaker's. Moreover, controlling the denoising strength in diffusion-based Adversarial Purification is crucial. As shown in Fig.\ref{fig:suc_sim}, with an increase in diffusion steps, although the success rate of the attack and the similarity to the target speaker decrease, the success rate of the defense and the similarity to the source speaker also eventually decrease after reaching a certain inflection point.

Additionally, it is important to note that the diffusion module can also introduce some damage to normal speech. As shown in Fig.\ref{fig:sim}, the similarity between the normal speech and the source speaker rapidly declines with increasing diffusion steps. This degradation can further affect the quality of TTS synthesis. Therefore, we introduced a ResNet18-based discriminator in front of the diffusion model. Experimental results show that, after training on both positive and negative samples, this discriminator achieved a 100\% recognition rate for adversarial samples of the attack types it was trained on.
We will evaluate unseen adversarial methods in the future.

\subsection{Zero-Shot TTS Synthesis Results}
We also explored the impact of different methods on the quality of zero-shot TTS synthesis. By extracting the speaker embeddings from adversarial samples and synthesizing speech, we evaluated the results by calculating the cosine similarity between the speaker embeddings of the synthesized speech and those of the source and target speakers. We observed that adversarial training defenses resulted in adversarial samples maintaining a high similarity to the source speaker, but with a relatively high similarity to the target speaker as well. In contrast, adversarial purification methods significantly reduced the similarity to the target speaker but also degraded a substantial portion of the source speaker’s information.

\section{CONCLUSION}
This paper explores adversarial attacks and robust defenses in speaker embedding based zero-shot TTS systems, including both traditional and LLM-based. In the adversarial attack, we employ PGD and Adam-based white-box attack methods to target the speaker encoder of the zero-shot TTS system, aiming to guide the TTS system into synthesizing speech that closely resembles the target speaker. To mitigate the potential threats these attacks posed, we implemented traditional active defense strategies, such as adversarial training, and novel passive defense strategies based on diffusion models for adversarial purification. We assessed the effectiveness of these defenses, their impact on model performance, and their effects on synthesis quality.

\section{Acknowledgement}
This research is funded in part by the National Natural Science Foundation of China (62171207), Guangdong Science and Technology Plan (2023A1111120012) and OPPO. Many thanks for the computational resource provided by the Advanced Computing East China Sub-Center.

\bibliographystyle{IEEEbib}
\bibliography{icme2025}

\end{document}